\documentclass[%
 reprint,
 superscriptaddress,
amsmath,amssymb,
 prx,
floatfix,
]{revtex4-2}

\usepackage{graphicx}
\usepackage{xcolor}

\usepackage{lineno}

\begin{document}

\title{Ultralow-Energy Measurements using the Startpoint of $\beta$ Decays} 

\author{D.~Kodroff}\email{danielkodroff@lbl.gov}
\affiliation{Lawrence Berkeley National Laboratory (LBNL), Berkeley, CA 94720-8099, USA}

\author{M.R.~Williams}\email{michaelwilliams@lbl.gov}
\affiliation{Lawrence Berkeley National Laboratory (LBNL), Berkeley, CA 94720-8099, USA}

\begin{abstract}
We propose a novel measurement of $\beta$ decays using low-temperature solid-state detector technologies. The $\beta$ startpoint, where the $\beta$ kinetic energy is zero, offers a unique probe of weak nuclear physics that has not yet been exploited experimentally. We describe how this technique enables searches for heavy sterile neutrinos in the keV--MeV energy range and show that, with current technologies, sensitivities to sterile neutrino coupling to electrons can be achieved that exceed the current best constraints. The spectrum produced by the recoiling daughter ion also offers a calibration of the nuclear recoil response, critical for both sub-GeV dark matter searches and coherent elastic neutrino–nucleus scattering (CE$\nu$NS) measurements. We outline a potential experimental scheme using neutron activation to produce $^{32}$P \textit{in situ} in a silicon substrate, discuss projected sensitivities to sterile neutrinos, and evaluate the prospects for nuclear recoil calibrations.
\end{abstract}

\maketitle

\section{Introduction}
\label{sec:intro}

Low-temperature solid-state detector technologies have reached unparalleled sensitivities to low-energy depositions~\cite{TwoChannelLimits,CPDLimits}. Transition Edge Sensors (TESs)~\cite{TwoChannelLimits,armatol2026lowenergyphononbursts,finkPerformanceLargeArea2021}, Neutron Transmutation-Doped thermistors (NTDs)~\cite{HALLER1994127,Det_review}, Kinetic Inductance Detectors (KIDs)~\cite{temples2025developmentstatuskipmdetector,MKIDPaper}, Metallic Magnetic Calorimeters (MMCs)~\cite{MMC2025}, and qubit-based sensors~\cite{qpd2025,squat2026} have each been demonstrated as viable technologies for scalable detector systems. This combination of low-energy thresholds and scalability has enabled nuclear recoil searches for dark matter~\cite{TwoChannelLimits,CPDLimits,CRESST2023Limits,MMC2025,EDELWEISS_TES,COSINUS}, low-energy photon detection for axions~\cite{squat2024,Abdelhameed:2020hys,Mirzakhani:2025bqz,Alloni:2026jgy}, calorimetric searches for neutrinoless~\cite{amore2025,CUPID:2025avs,Auguste:2026pey,Auguste:2026pey} and two-neutrino~\cite{CUPID-Mo,PhysRevLett.131.222501,CUORE} double-$\beta$ decay, detection of coherent elastic neutrino–nucleus scattering (CE$\nu$NS)~\cite{NUCLEUS,RICOCHET,MINER:2025wrk}, sterile neutrino searches via $\beta$ spectrometry~\cite{Lee:2024pml,Lee:2025txi,Lee:2026tzu} and electron capture decays~\cite{beest2022,beest2025,holmes2025,echo2025}. In this article, we propose a novel measurement of $\beta$ decays in the limit of the $\beta$ particle being emitted at rest to search for hypothetical sterile neutrinos. 

In a Standard Model (SM) $\beta$ decay, the $\beta$ energy spectrum is determined by the kinematics of a three-body decay in which the emitted neutrino carries away a variable fraction of the total energy and momentum, resulting in a continuous spectrum. Much of the experimental effort aimed at probing neutrino properties has focused on the region near the $\beta$ endpoint, where the kinetic energy of the emitted $\beta$, $T_\beta$, approaches the $Q$-value of the decay, $Q$. The full $\beta$ and two-neutrino double-$\beta$ spectral shapes, including their low-energy portions, are also actively measured~\cite{In115-a,ACCESS:2023gdy,113Cd,CUPID-Mo,CUORE}. Here we focus instead on the $\beta$ startpoint, where the $\beta$ kinetic energy goes to zero ($T_\beta \to 0$), which corresponds to the maximum neutrino energy. At this point in the spectrum, the emitted $\beta$ is at rest, such that the emitted neutrino and recoiling daughter nucleus emerge back-to-back. If such a decay occurs within the substrate of a solid-state detector, the nuclear recoil of the daughter ion can be measured as athermal phonons, while the neutrino escapes the detector, as illustrated in Fig.~\ref{fig:detscheme}. We show in this work that the $\beta$ startpoint spectrum produces a distinctive spectral shape that offers a unique probe of beyond-the-Standard-Model (BSM) weak nuclear physics, such as the emission of sterile neutrinos.

The non-zero neutrino mass necessitates extensions to the SM, the simplest of which introduces massive right-handed neutrinos, so-called sterile neutrinos, which mix with active neutrinos~\cite{Gouvea_2016}. In the minimal (3+1) extension, the electron flavor state is a superposition of mass eigenstates, $\nu_e = \sum_{i=1}^{4} U_{ei}\,\nu_i$, where $U_{ei}$ are elements of the extended Pontecorvo--Maki--Nakagawa--Sakata matrix and $\nu_4$, with mass $m_4$, is predominantly sterile. A fraction $|U_{e4}|^2$ of $\beta$ decays with $m_4 < Q$ then proceeds via emission of $\nu_4$. Sterile neutrinos spanning sub-eV to GeV masses are well motivated, with searches for these particles being a high priority in the neutrino community~\cite{Giunti:2019aiy,Dasgupta:2021ies}. Sterile neutrinos in the keV mass range are also compelling dark matter candidates~\cite{Boyarsky:2018tvu}. 

The distinct spectral shape of the $\beta$ startpoint is sensitive to the emission of massive neutrinos. We propose that a search for sterile neutrinos using this method complements other ongoing and proposed searches, including measurements near the $\beta$ endpoint of $^{3}$H decays~\cite{KATRIN:2026ywk}, kinematic reconstruction of $\beta$ decays~\cite{Carney:2022pku}, and measurements of electron capture decays~\cite{beest2022,holmes2025,echo2025}. By exploiting the spectral shape of $\beta$ decays near the startpoint, we demonstrate that with current detector technologies, sensitivities to the sterile neutrino coupling to electrons, $|U_{e4}|^{2}$, in excess of current best constraints can be achieved.

The recoil spectrum of the daughter nucleus also provides an opportunity to calibrate the nuclear recoil response of solid-state detectors: an essential task for both sub-GeV dark matter and CE$\nu$NS experiments utilizing phonon sensors. These searches typically rely on photon calibrations to determine the energy response and resolution of the detectors, under the assumption that the detector response to electron recoils and nuclear recoils is equivalent. However, this assumption has been challenged by experiments~\cite{beest_2021,CRESST2023Limits,CRESST:2024cpr} that have observed energy resolutions far worse than predicted from photon calibrations. Measuring the recoil spectrum of the daughter nucleus offers a direct and robust method for calibrating the nuclear recoil response at the lowest recoil energies achieved to date.

This article describes how the characterization of the $\beta$ startpoint opens avenues for multiple novel low-energy measurements. The text is organized as follows. Sec.~\ref{sec:form} begins with a definition of the $\beta$ startpoint and its predicted spectral shape. In Sec.~\ref{sec:exp}, we discuss the proposed experimental scheme and how it may be achieved. Sec.~\ref{sec:Sens} presents projected sterile neutrino sensitivities, and Sec.~\ref{sec:cal} discusses the calibration opportunities using the aforementioned experimental scheme. Sec.~\ref{sec:discussion} delves into additional systematic effects and paths forward, and we conclude with the prospects in Sec.~\ref{sec:outlook}.

\section{Formalism}
\label{sec:form}
\subsection{Sensitivity to non-zero neutrino mass}

The $Q$-value of a $\beta$ decay defines the kinetic energy, $T$, available to the progeny isotopes and the nonzero neutrino mass, $m_{\nu}$. The canonical $Q$-value is defined by the difference in masses of the parent and progeny isotopes, factoring out the rest mass of the electron:
\begin{equation}
\label{eq1}
    Q = T_{\nu} + m_{\nu} + T_{d} + T_{\beta}~\text{,}
\end{equation}
where $T_{\nu}$, $T_{d}$, and $T_{\beta}$ are the kinetic energies of the emitted neutrino, recoiling daughter ion, and emitted $\beta$, respectively. Here $m_\nu$ denotes the mass of the emitted neutrino mass eigenstate, with $m_\nu \approx 0$ for the light states and $m_\nu = m_4$ for sterile neutrino emission. We define the startpoint of the $\beta$ spectrum as: \begin{equation}
\label{eq2}
\begin{aligned}
    T_{\beta} &\rightarrow 0 ~\text{,} \\
    p_{\nu} &= p_{d} ~\text{,}\\
\end{aligned}
\end{equation} 
where $p_{\nu}$ and $p_{d}$ are the momenta of the emitted neutrino and recoiling daughter ion, respectively.

First, the energy of the emitted $\beta$ approaches zero, i.e. it is emitted at rest. This can be contrasted with the $\beta$ endpoint at which $T_{\nu} \rightarrow 0$. Second, the emitted neutrino and recoiling daughter ion must be back-to-back such that their momenta are equal.

Given these conditions, the kinetic energy of the recoiling daughter ion at the $\beta$ startpoint is 
\begin{equation}
\label{eq3}
    T_{d} = Q - E_{\nu} = Q + m_{d} - \sqrt{m_{d}^{2} + m_{\nu}^{2} + 2m_{d}Q} ~\text{.}  
\end{equation}
The dependence on the $Q$-value and $m_{d}$ serves to constrain which $\beta$-decaying isotopes are experimentally feasible. For example, $^{3}$H $\beta$ decays ($Q=18.59$~keV) have a recoiling ion energy of $T_{d}=61.5$~meV at the $\beta$ startpoint. Such a small energy is not currently measurable with existing technologies. In contrast, more energetic $\beta$ decays from $^{32}$P ($Q=1710.66$~keV) and $^{90}$Y ($Q=2278.5$~keV) have $T_{d}=49.13$~eV and $T_{d}=31.00$~eV, respectively, at the $\beta$ startpoint.  These isotopes provide an accessible benchmark which can be measured with existing technologies. Note that the $Q$-value is known to a precision of 40~eV for $^{32}$P and 160~eV for $^{90}$Y~\cite{Wang_2021}. These uncertainties lead to a systematic uncertainty in $T_{d}$ at the $\beta$ startpoint of 2~meV and 12~meV, respectively.

\begin{figure}[t]
	\centering
	\includegraphics[width=0.99\columnwidth]{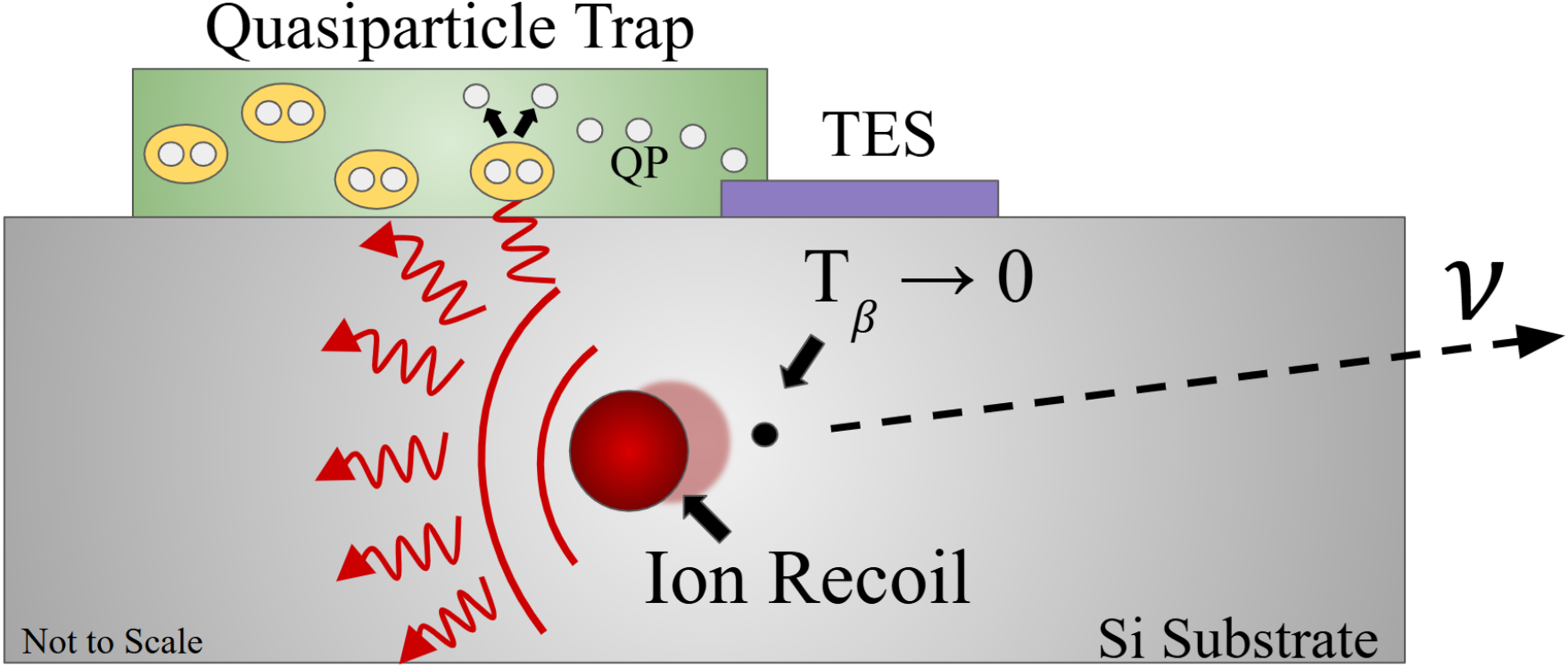}
	\caption{Schematic of the detection principle of the $\beta$ startpoint using solid-state detectors. Near the $\beta$ startpoint, the $\beta$ has near zero kinetic energy such that the recoiling daughter ion and neutrino are approximately back-to-back. The observable energy deposited in the silicon substrate is that of the recoiling ion plus the small kinetic energy from the $\beta$. The kinetic energy produces athermal phonons which break Cooper pairs in a quasiparticle trap which tunnel into the active transition edge sensors, giving the observed signal.}
	\label{fig:detscheme}
\end{figure}

The measured energy in a sensitive athermal phonon detector, like that shown in Fig.~\ref{fig:detscheme}, would be the sum of the recoiling ion energy and the $\beta$ energy ($T_d +T_\beta$), assuming the neutrino leaves the target volume without interacting. The predicted observed spectrum for $^{32}$P in such a detector, considering a 1~eV detector resolution, is shown in Fig.~\ref{fig:P32spec}. The spectrum exhibits a shoulder where the minimum energy corresponds to the $\beta$ startpoint, and the higher energies are those where the emitted $\beta$ has increasingly larger kinetic energies. 

The ion recoil energy at the $\beta$ startpoint provides sensitivity to the neutrino mass, as shown in Eq.~\ref{eq3}. The amount by which the recoiling ion energy changes as a function of neutrino mass can be calculated by expanding Eq.~\ref{eq3} in powers of $m_{\nu}$
\begin{equation}
\label{eq4}
    \Delta T_{d} = -\frac{m_{\nu}^{2}}{2\sqrt{m_{d}^{2} + 2m_{d}Q}} + O(m_{\nu}^4) + ... 
\end{equation}
The contribution from $O(m_{\nu}^4)$ and higher terms is greatly suppressed with respect to the leading order term and can be safely neglected. A full derivation of the above calculations can be found in the Supplemental Materials.

The purple, blue, and pink lines in Fig.~\ref{fig:P32spec} show the predicted spectra for sterile neutrino masses $m_4 = 500$, 1000, and 1500~keV, respectively, with coupling strength $|U_{e4}|^{2}=10^{-3}$. This benchmark is chosen for illustrative purposes only, as it lies near the existing experimental constraints (Fig.~\ref{fig:sensitivity}). The emission of these heavy neutrinos shifts the $\beta$ startpoint to lower energies by 4.2~eV, 16.8~eV, and 37.8~eV, respectively. 

The quadratic dependence on neutrino mass in Eq.~\ref{eq4} limits the ability of $\beta$ startpoint measurements to perform neutrino mass measurements. Neutrino masses on the meV-scale would produce vanishingly small shifts to the $\beta$ startpoint energy. However, massive neutrinos on the order of 100~keV or greater would lead to $\Delta T_{d}$ values measurable by current or near-future experiments. In this mass regime, sterile neutrinos may also be a viable dark matter candidate~\cite{Boyarsky:2018tvu}.

\subsection{Effect on $\beta$ Spectral Shape}

The emission of a heavy sterile neutrino also necessarily affects the $\beta$ kinetic energy distribution. As the sterile neutrino mass is increased, a larger fraction of the $Q$-value energy is apportioned to $m_{4}$ as opposed to $T_{\beta}$. The resulting spectral shape for $T_{\beta}$ will be altered in addition to being shifted to lower energies. The change in spectral shape depends on the isotope. If we consider, for example, the Gamow-Teller allowed $^{32}$P decay, the phase-space factor depends on neutrino mass as
\begin{equation}
\label{eq5}
    f ~ \propto ~ p_{\beta} \cdot E_{\beta} \cdot (Q-T_{\beta}) \cdot \sqrt{(Q-T_{\beta})^{2} - m_{4}^{2}} ~ \text{,}
\end{equation}
where $E_{\beta} = T_{\beta}+m_{\beta}$ is the total electron energy and $p_{\beta}$ is the momentum of the electron. 

\begin{figure}[!t]
	\centering
	\includegraphics[width=0.99\columnwidth]{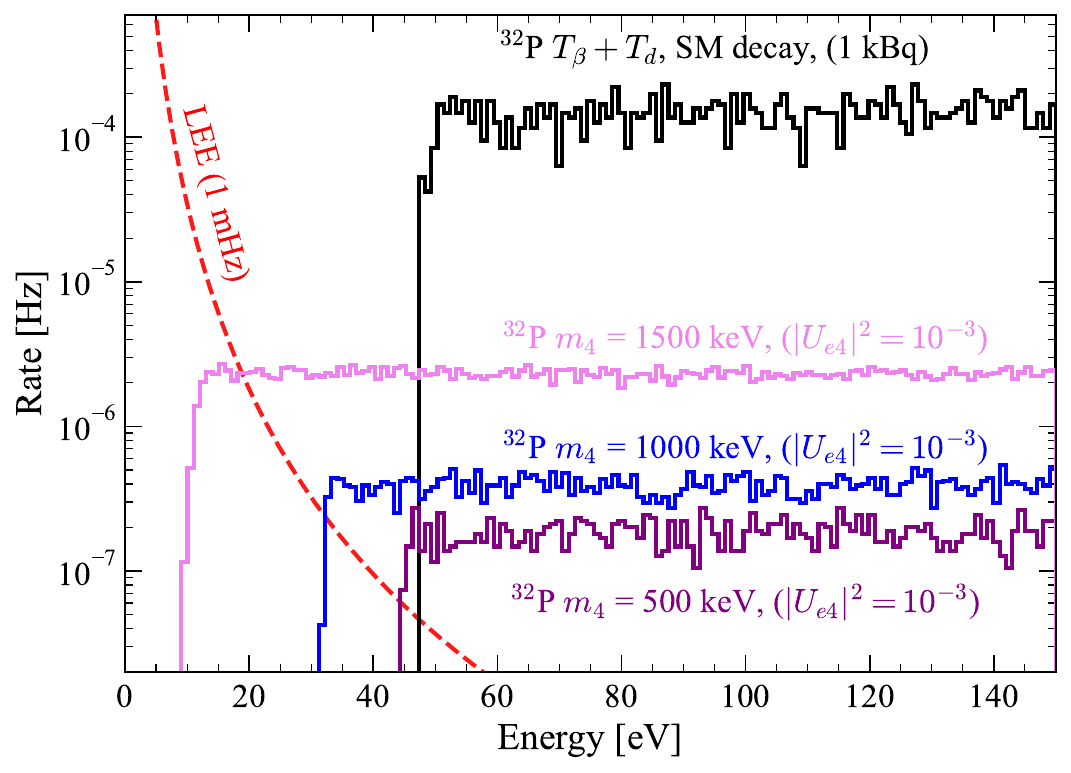}
	\caption{The predicted $^{32}$P $\beta$ startpoint spectrum generated using Monte Carlo simulations with an assumed 1~eV detector resolution. The predicted Standard Model decay (black) is the sum of the recoiling daughter $T_{d}$ and emitted $\beta$ $T_{\beta}$ with an assumed activity of 1~kBq. The roll-off is predicted at the $\beta$ startpoint of 49.1~eV. Decays in which a sterile neutrino with $m_4 = 500$~keV, 1000~keV, or 1500~keV is emitted, with a coupling $|U_{e4}|^{2}=10^{-3}$, are shown in purple, blue, and pink, respectively. The emission of the massive neutrino shifts the $\beta$ startpoint by 4.2~eV, 16.8~eV, and 37.8~eV, respectively. Also shown is an example low-energy excess (LEE) background power-law assuming a 1~mHz rate and shape as from~\cite{armatol2026lowenergyphononbursts}.}
	\label{fig:P32spec}
\end{figure}

Not shown in Eq.~\ref{eq5} are the contributions from the Fermi function and additional shape corrections for non-allowed $\beta$ decay types. The latter term is relevant for the unique first-forbidden $\beta$ decay of $^{90}$Y. This amended phase space factor is accounted for in the Monte Carlo simulation results shown in Fig.~\ref{fig:P32spec} and used throughout this work. 

\section{Experimental Scheme}
\label{sec:exp}

A schematic of a potential experiment to measure the $\beta$ startpoint can be seen in Fig.~\ref{fig:detscheme}. Our starting point is a sensor like that discussed in~\cite{TwoChannelLimits}: a 1~cm~$\times$~1~cm~$\times$~0.1~cm silicon crystal with two channels of athermal phonon sensors based on the quasiparticle trap architecture~\cite{irwinQuasiparticleTrapAssisted1995}. The substrate is assumed to be natural silicon with ppm-level dopants (i.e. $^{31}$P). This technology has achieved energy resolutions of $O(100~\text{meV})$~\cite{TwoChannelLimits}. However, these detectors observe an unknown low-energy background, the so-called low-energy excess (LEE), that is well described by a power-law, but whose rate can vary from run-to-run~\cite{Baxter_2025}. 

In our benchmark scenario, we target a 1~kBq activity of our $\beta$ decaying isotope within the bulk silicon. Thus, we assume the silicon to be well described as naturally abundant silicon with ppm-level impurities. This activity is chosen to optimize the event rate without suffering appreciable event pileup effects in these detectors~\cite{TwoChannelLimits}. The devices of Ref.~\cite{TwoChannelLimits} exhibit pulse rise times of $O(10\,\mu\mathrm{s})$ and fall times of $O(100\,\mu\mathrm{s})$, so that at 1~kBq, corresponding to a mean interval between decays of 1~ms,  the fraction of events affected by pileup remains manageable.

It is worth quantifying the level of doping implied by this scheme. The devices of Ref.~\cite{TwoChannelLimits} use float-zone silicon with residual dopant concentrations below about 10$^{12}/\text{cm}^3$, whereas the wafer proposed here contains a $^{31}$P density of 10$^{17}/\text{cm}^3$ (approximately 2~ppm). For the phonon channel we expect this to be tolerable. Phonon transport in natural silicon is dominated by two processes: anharmonic decay of high-frequency phonons, and scattering of phonons off isotopic mass defects, whose density exceeds that of substitutional phosphorus by four orders of magnitude. In addition, the donors are frozen out at mK temperatures. Demonstrating that the $O$(100~meV) resolution of Ref.~\cite{TwoChannelLimits} survives ppm-level doping is nevertheless an essential  step of the R\&D program. Thus in a purely athermal phonon sensor, we believe this level of dopant is workable.

We first consider the ideal $\beta$-decaying isotope with which to perform measurements of the $\beta$ startpoint. The ideal isotope should entirely decay to the ground-state, be measured assuming demonstrated detector energy resolutions of $O(100~\text{meV})$, and be well separated from the previously observed LEE spectra. This leads to $^{32}$P and $^{90}$Y as potential candidates. Yet, $^{32}$P is closer in atomic mass to silicon than $^{90}$Y, so it offers a better kinematic match for a measurement of energy resolution to nuclear recoils at the $\beta$ startpoint. Production of $^{32}$P can also be reasonably performed using a variety of techniques. 

We consider three methods for implanting $^{32}$P into the silicon wafer: (1) embedding the radioactive isotopes directly into the silicon at an accelerator, (2) diffusing the radioactive isotope into the silicon, and (3) producing the radioactive isotope \textit{in situ} using neutron activation. 

Method (1) is likely achievable with existing technologies and facilities such as Isotope Separation On-Line (ISOL) at TRIUMF~\cite{Kunz:2023oqv,Mostamand:2020woh} or in-flight fragmentation at FRIB~\cite{Balantekin:2014opa}. However, at the typical implantation energies of $O$(10~keV), the $^{32}$P isotopes would be placed only in the first $\sim$100~nm of the silicon surface. For the uniform embedding in a 0.1~cm thick device desired in this study, reaching deeper into the substrate would require MeV implantation energies. While this is experimentally feasible~\cite{Wenander:2013pjd,Schmidt:2018gle}, the large fluence of $^{32}$P would, in either implantation case, generate substantial crystal lattice damage in the form of vacancies, interstitials, and cluster defects~\cite{Pintiliie:2009njs}. These defects have been hypothesized to be a source of the LEE~\cite{DefectLEE} and have been shown to lead to elevated event rates~\cite{armatol2026lowenergyphononbursts}. Thermal annealing has been shown to repair the silicon crystal lattice via defect migration and vacancy-interstitial recombination~\cite{ccdanneal1,ccdanneal2}, though it is not clear if all damage can be repaired. However, thermal annealing may adversely affect the metal quasiparticle trap films. Above $300^{\circ}$~C, aluminum films see hillock formation and eventual melting~\cite{MARTIN199564}, while the superconducting transition temperature ($T_c$) of tungsten has been known to shift when exposed to high temperatures~\cite{KAIDATZIS201661}. Any post-implantation annealing time would additionally deplete the $^{32}$P activity given its 14.3~d half-life. The implantation, annealing, and data collection would have to occur within the first two half-lives. While this is certainly possible, the potential silicon lattice damage, involved procedural steps, time pressure, and risk of radioactive exposure during this process lead us to believe this is not the ideal choice.

Method (2) is also achievable as phosphorus is one of the most extensively studied dopants in silicon~\cite{2022SEMSC.23411410D,2004ApPhL..85.1359O,2002ApPhL..81.3197O}. Indeed, radioactive $^{32}$P-doped oxide P$_{2}$O$_{5}$ has previously been diffused into SiO$_{2}$ at 1100$^\circ$~C~\cite{GHOSHTAGORE1975399}. The typical approach to the diffusion process involves exposing the surface of a silicon substrate wafer to a phosphorus-bearing source — either phosphine ($^{31}$PH$_{3}$) gas, POCl$_{3}$ vapor, or a deposited phosphosilicate glass (PSG) layer — at temperatures of 800–1200$^\circ$~C in a furnace tube, followed by higher temperature annealing. At high temperatures, phosphorus atoms diffuse through the silicon lattice primarily via an interstitial-substitutional mechanism, in which silicon self-interstitials exchange with substitutional phosphorus atoms, enabling atomic transport. The diffusion of phosphorus in silicon can to first order be described by the Arrhenius relation $D = D_{0} e^{\frac{-E_{a}}{kT}} $, where $E_{a}$ is the the activation energy for phosphorus diffusion in silicon at 3.7~eV~\cite{sze_physics_2007}, $D_{0}\sim3.85$~cm$^{2}$/s~\cite{Fair_1977}, $k$ is the Boltzmann constant and $T$ the temperature. With 10 hours at 1200$^\circ$~C, diffusion lengths of $\sim2~\mu$m can be achieved. Method (2) is somewhat preferable to Method (1): larger depths can be achieved and the formation of large defect complexes in the crystal lattice may be mitigated. High temperature treatments of silicon substrates have also shown the ability to remove other impurities, such as radioactive tritium~\cite{Saldanha:2025ayz}. 

This method is not without its own challenges. For this process to be feasible, a commercially available aqueous $^{32}$P source would have to be converted into a surface-stable diffusion source (akin to PSG) that can be placed on the silicon substrate before loading into a furnace. These steps would most likely have to occur prior to the deposition of the thin films (e.g. aluminum and tungsten) required to form the active sensors due to the temperature sensitivities of the films discussed earlier. Depositing the films following the injection of $^{32}$P would pose a radiation hazard during handling and production of the thin films. Despite the plausibility of this method, experimental challenges remain to safely perform this diffusion into ultra-pure semiconductor-grade silicon with a large activity of $^{32}$P without any additional impurities.

We consider Method (3) as the simplest route to observing the $\beta$ startpoint with $^{32}$P. Neutron activation leverages the fact that n-type semiconductors doped with $^{31}$P are readily available. Typically, n-type semiconductors with $^{31}$P are produced via neutron transmutation on pure silicon~\cite{IAEA_NTD_2012}: thermal neutron capture, $(n,\gamma)$ on $^{30}$Si, produces $^{31}$Si, which decays with a 2.62~h half-life to stable $^{31}$P. Thus, the same thermal neutron capture process can be exploited to produce $^{32}$P using commercially procured $^{31}$P-doped wafers as a starting point. A key benefit of this technique is that the isotopes of interest are produced within the existing silicon lattice structure and are distributed uniformly. Thermal neutron interactions may also lead to defect formation, though this has not yet been demonstrated in cryogenic detectors. We note that measurements in which the recoil energy following neutron capture is known \textit{a priori} have been proposed as a direct probe of defect creation and of the associated energy loss in cryogenic absorbers~\cite{Crab_defect}. Such measurements would provide input complementary to the quenching study described in Sec.~\ref{sec:discussion}.

To produce the required 1~kBq activity of $^{32}$P in a 0.1~cm$^{3}$ silicon wafer via neutron capture, sufficiently large neutron fluxes are required. Nuclear reactors are typically able to achieve thermal neutron fluxes in excess of 10$^{13}$~n/cm$^{2}$/s~\cite{BORIODITIGLIOLE2014249}. Assuming a thermal neutron flux described by a Gaussian ($\mu=0.025$~eV, $\sigma=0.5$~eV) with a total flux of 10$^{13}$~n/cm$^{2}$/s, the desired $^{32}$P activity can be achieved with about 2 days of exposure. This would require starting with $\sim10^{16}$~$^{31}$P atoms in the 0.1~cm$^{3}$ wafer, corresponding to an n-type silicon wafer of $\sim0.1~\Omega\cdot$cm resistivity.

As a result of this process, radioactive isotopes with half-lives longer than $^{32}$P are also produced. These are $^{32}$Si and $^{33}$P, whose half-lives are 153~y and 25.4~d, respectively. The $Q$-values of these $\beta$ decays are 227.3~keV and 248.5~keV, corresponding to $\beta$ startpoints of $0.86$~eV and $1.0$~eV, respectively. Thus, they would manifest as effectively flat backgrounds in the actual measurement. However, the activities of these isotopes are suppressed by approximately 11 and 7 orders of magnitude, respectively, with respect to that of $^{32}$P, placing them well below the predicted LEE and signal rates. See further discussion in the Supplemental Materials.

\section{Sterile Neutrino Sensitivity}
\label{sec:Sens}

Here we present projected sensitivities assuming $^{32}$P is present at an activity of 1~kBq within an athermal phonon silicon detector. Figure~\ref{fig:P32spec} shows the predicted spectra for the $^{32}$P $\beta$ startpoint and higher energy spectrum in black. The spectrum was produced via Monte Carlo simulations of the $^{32}$P $\beta$ decay spectrum in \textsc{GEANT4}~\cite{AGOSTINELLI2003250} smeared with detector energy resolution applied. The analytically predicted $\beta$ startpoint energy of 49.1~eV is recovered. The LEE spectrum (dashed red) corresponds to a 1~mHz rate and is modeled by a power-law fit with slope corresponding to the spectrum in Ref.~\cite{armatol2026lowenergyphononbursts}. Also shown are three example spectra assuming the presence of a sterile neutrino. Here we assume a coupling constant of $|U_{e4}|^2=10^{-3}$. In these projections, we only consider the Standard Model $^{32}$P $\beta$ decays and the LEE as backgrounds.

\begin{figure}[!t]
	\centering
	\includegraphics[width=0.99\columnwidth]{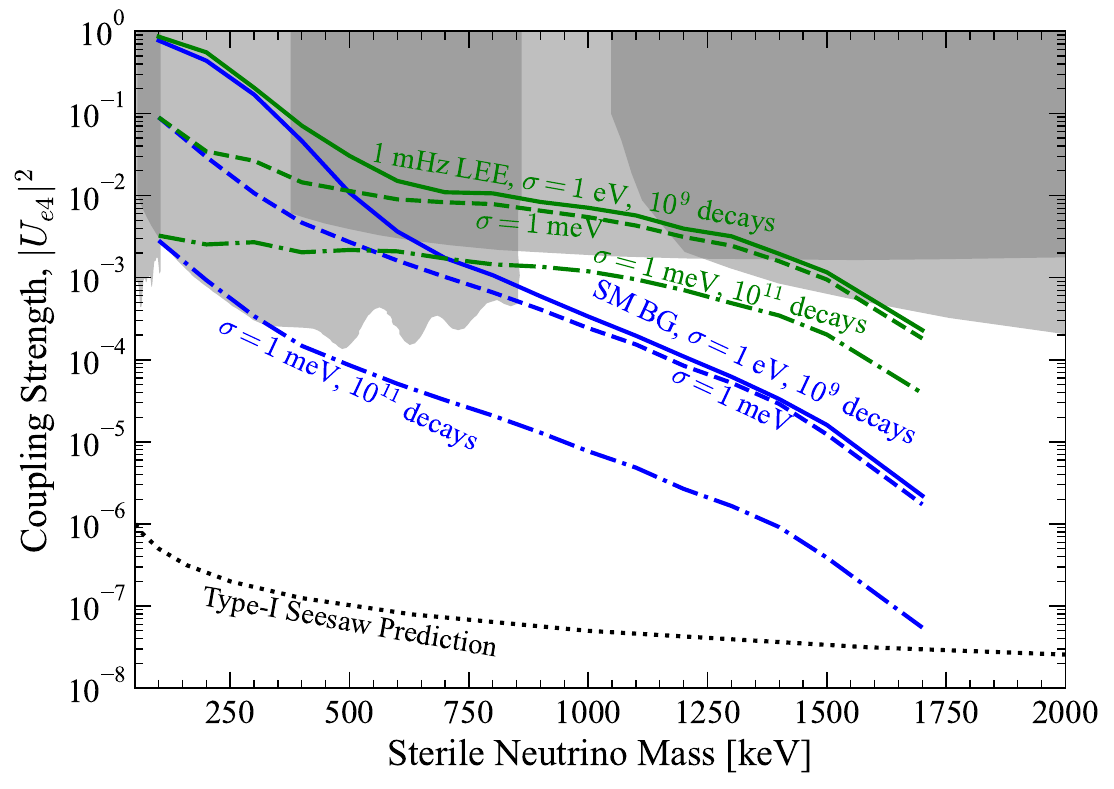}
	\caption{Projected sensitivity (90\% confidence level) to sterile neutrinos using the Yellin optimum interval technique~\cite{Yellin}. The projections in blue are those in which only $^{32}$P SM decays are considered as a background. The green lines additionally consider a 1~mHz low-energy excess (LEE) background component as discussed in the text. The solid(dashed) lines correspond to an applied detector energy resolution of 1~eV(1~meV). The solid and dashed lines assume a 1~kBq activity of $^{32}$P with a 30~d exposure. The dash-dotted line shows projections with a factor of 100 increase in counts for both background assumptions given a 1~meV energy resolution. In the most optimistic scenario, in which no LEE background is present, sensitivity to the Type-I seesaw prediction can be approached~\cite{Bolton:2019pcu}. Exclusions from existing experiments are also shown in shaded gray~\cite{calaprice_f20,beest_2021,Borexino_2013}.}
	\label{fig:sensitivity}
\end{figure}

Figure~\ref{fig:sensitivity} shows the 90\% confidence levels on the projected sensitivity using the Yellin optimum interval method~\cite{Yellin}: a technique typically used within the direct detection dark matter community to handle an unmodeled LEE background that can vary from detector to detector. The blue lines show the projected sensitivities assuming no LEE background is present, whereas the green lines show the sensitivity in the presence of the LEE background. In both cases the SM $^{32}$P decays are considered. We further show the impact of both 1~eV (solid) and 1~meV (dashed) energy resolution. At the assumed 1~kBq activity there are 1.8$\times10^{9}$ $^{32}$P atoms in the 0.1~cm$^{3}$ silicon substrate. We assume continuous data taking for a one month period in which, on average, there will be 1.4$\times10^{9}$ decays. This exposure is shown in the solid and dashed lines. 

The dash-dotted lines show a two order of magnitude increase in exposure. This could be achieved in approximately one calendar year if 10 detectors were operated in parallel for a month, for 10 months in a row, allowing time for cooldown and warmup cycles. Such a multi-detector campaign is consistent with the scalability goals of athermal phonon detector efforts. Beyond increased exposure, improvements in detector energy resolution from the current $O(1$~eV) toward the $O(1$~meV) scale would substantially enhance the sensitivity to the lowest sterile neutrino masses where the startpoint shift is smallest. 

At low sterile neutrino masses, the sensitivity is limited by the detector energy resolution such that the SM $^{32}$P decays are the primary background. Equation~\ref{eq4} shows that smaller neutrino masses lead to smaller shifts in the $\beta$ startpoint, which require improved detectors to resolve the sterile neutrino contribution from the SM $\beta$ startpoint. The sensitivity improves as the sterile neutrino mass increases due to the sterile neutrino spectrum shifting further from the $\beta$ startpoint and the increased fraction of decays in this low-energy regime. At larger sterile neutrino masses, the LEE background becomes the limiting factor.

Existing experimental constraints are shown as shaded gray regions. The current best constraints in this parameter space are from the BeEST collaboration using electron capture decays of $^{7}$Be below its $Q$-value of 862~keV~\cite{beest_2021}, analysis of the $^{20}$F $\beta$ decay spectrum~\cite{calaprice_f20}, and measurements of sterile neutrino mixing using $^{8}$B solar neutrinos observed in Borexino~\cite{Borexino_2013}.

These projections show that even in the presence of the LEE background, new parameter space in the 850 keV to 1700 keV mass range can be excluded. In the optimistic scenario of no LEE backgrounds, the projected sensitivity approaches to within approximately a factor of two of the benchmark prediction of the Type-I seesaw mechanism, $|U_{e4}|^2 \sim m_{\nu}/m_{4}$ assuming a light neutrino mass $m_\nu = 0.05$~eV~\cite{Bolton:2019pcu}, at the upper end of the mass range considered.

The LEE background is the dominant background with the largest systematic uncertainty at low energies. The spectral shape and normalization of the LEE have been observed to vary between detectors and between cooldowns of the same detector~\cite{angloherLatestObservationsLow2022}. In the projections presented in Fig.~\ref{fig:sensitivity}, we have shown the impact of a power-law LEE model with a rate of 1~mHz. If the LEE can be mitigated or removed, an active area of R\&D within the cryogenic detector community, the sterile neutrino sensitivity would improve substantially, shifting closer to the background-free projections. Conversely, if the LEE rate is higher than assumed, the sensitivity at large sterile neutrino masses would be degraded. We note that if the LEE background is sufficiently well characterized in the future, more powerful statistical methods such as profile likelihood analyses could be employed, further improving the sensitivity and enabling a quantitative assessment of discovery potential.

Beyond the dominant LEE and Standard Model $^{32}$P $\beta$ decays, neutron-activation products with half-lives exceeding that of $^{32}$P, namely $^{32}$Si and $^{33}$P, can be produced, but their activities are suppressed by $\sim$11 and $\sim$7 orders of magnitude, respectively (see the Supplemental Materials). Their low $Q$-values render their spectra effectively flat in Fig.~\ref{fig:P32spec}, at rates far below the scale shown. Trace isotopes from cosmogenic and fast-neutron activation, such as $^{3}$H and $^{22}$Na~\cite{Saldanha2020}, may also be produced. All of these contributions are expected to be subdominant to the LEE and are therefore excluded from the sensitivity projections; should the LEE be suppressed or eliminated, they would need to be reconsidered.

\section{Nuclear Recoil Calibration}
\label{sec:cal}

Measurements of the $\beta$ startpoint also introduce a potential nuclear recoil calibration method for low-energy threshold detectors. Athermal phonon sensors are typically calibrated with eV-scale photons produced by a laser diode and injected via optical fibers. Existing detectors have demonstrated the ability to clearly observe single photons that deposit energy into the silicon substrate with energy resolutions nearing 200~meV~\cite{TwoChannelPaper}. We note that photon calibrations using energies greater than the bandgap of the substrate produce largely surface interactions ($<100$~nm), whereas one searches for recoils produced uniformly in the bulk substrate. A key assumption for dedicated athermal phonon searches for dark matter is that if no ionization signal is collected, all the quanta produced in an interaction \textit{should} eventually go into the phonon channel. This leads to the assumption that the energy resolution for nuclear recoils is identical to that of electronic recoils. Yet, no such \textit{in situ} nuclear recoil calibration has been performed in athermal phonon detectors at energies below 100~eV to confirm this assumption in the relevant low-energy regime. 

The observation of the $\beta$ startpoint feature enables a direct probe of the nuclear recoil response in athermal phonon detectors. This allows us to address two important questions: (1) what is the scale of defect-induced quenching for low-energy nuclear recoils, and (2) is there variation in the detector energy resolution between nuclear and electronic recoils?

The readily predictable ion recoil energy and spectral shape of the $^{32}$P $\beta$ decay allow both questions to be addressed directly. While the spectrum is predicted to have a clear startpoint at 49.1~eV, if nuclear recoils caused by the daughter ion lose some of their energy to the creation of Frenkel defects, then the spectrum would instead be shifted to lower energy by a constant value, corresponding to the Frenkel pair defect creation energy ($\sim$~25~eV in silicon~\cite{defectenergy}). The ramifications of such an observation are broad, and would require the correction of any previous dark matter search that assumed equivalence between electronic recoil (ERs) and nuclear recoil (NRs). Second, given that photon calibration sources are typically tunable, a measurement of the absolute athermal phonon yields can be done at the same energy for both nuclear and electron recoils for the first time. The nuclear recoil resolution can be measured simultaneously by fitting the onset of the startpoint spectrum with the predicted spectrum convolved with a detector response function (approximately an error function near the onset), analogous to Compton-edge fits, while the electron recoil resolution is obtained from in situ photon calibrations. Since only events near the onset constrain the width, this estimate is statistically less precise than a peak-based one, and it depends on the assumed response model: unmodeled broadening of the nuclear recoil response would be absorbed into the fitted resolution.

The presence of defect-induced quenching would shift the predicted sterile neutrino signals in Fig.~\ref{fig:P32spec} to lower energies. In turn, the sensitivity to sterile neutrinos would be limited by the increasing overlap with the LEE background. However, future improvements in detector resolution and the mitigation or removal of the LEE background — both of which are top R\&D priorities — would remedy this. 

\subsection{Phonon and Ionization Readout}
\label{sec:cal2}

The measurements described above rely solely on the athermal phonon signal from the recoiling daughter ion. However, detectors capable of simultaneously reading out both ionization and athermal phonon signals — such as those employed by SuperCDMS~\cite{cdmssensproj}, EDELWEISS~\cite{EDELWEISS_TES}, and proposed by TESSERACT~\cite{armatol2026lowenergyphononbursts} — could provide significant additional discriminating power in this measurement. 

In a semiconductor detector operated at cryogenic temperatures, an energy deposition produces both athermal phonons and electron-hole pairs. The ratio of the ionization yield to the total recoil energy will differ between ERs and NRs due to quenching for NRs: a phenomenon quantified by the Lindhard factor~\cite{Lindhard1961}. In the context of the present measurement, the $^{32}$P $\beta$ decay produces both an ER component (from the emitted $\beta$ particle) and an NR component (from the recoiling $^{32}$S daughter ion). Near the $\beta$ startpoint, where the $\beta$ kinetic energy approaches zero, the interaction is predominantly nuclear recoil in character, with a small ER admixture from the near-threshold $\beta$ electron. A 49.1~eV nuclear recoil produces at most one or two electron–hole pairs and therefore cannot be identified event by event at present charge-amplifier noise levels. A large sample of startpoint events nevertheless allows the mean ionization yield to be determined statistically, and single-charge-sensitive devices could directly measure the probability of producing at least one pair, probing the silicon ionization threshold.

The primary challenge of implementing dual-channel readout in this context is the requirement to collect charge from a device that also hosts the phosphorous. The electrode geometry must be designed such that the ionization signal can be drifted efficiently to the collection electrodes without recombination or trapping. SuperCDMS-style interleaved Z-sensitive ionization and phonon (iZIP) detectors have demonstrated high charge collection efficiency in silicon, and the iZIP geometry could in principle be adapted to the neutron-activated wafer geometry described in Sec.~\ref{sec:exp}. However, the dopant density and detector geometry, e.g. 0.1~cm thickness, would likely have to be amended for charge sensitive detectors as the phosphorous are likely to trap charge, quenching the ionization signal. Shorter drift lengths and lower dopant densities are likely required. The R\&D needs for this detector concept are beyond the scope of this work. Despite these challenges, a dual-channel readout represents another compelling path for future iterations of this measurement and is within the demonstrated capabilities of existing cryogenic detector technology.

\section{Discussion}
\label{sec:discussion}

We have presented the framework for a novel measurement of weak nuclear decays via the $\beta$ startpoint, enabled by advances in low-threshold solid-state detectors. This spectral measurement has the dual benefits of enabling searches for heavy sterile neutrinos and studies of the nuclear recoil response in athermal phonon detectors. In this section, we discuss the systematic effects, practical considerations, and broader implications of this technique. 

\subsection{Atomic Effects}

Beyond defect-induced quenching, additional atomic-scale processes could systematically affect the shape of the $\beta$ startpoint spectrum in ways that are challenging to predict from first principles. Effects can arise from the emitted $\beta$ (e.g., shake-off/shake-up processes and bremsstrahlung) or from the recoiling daughter nucleus (e.g., the Migdal effect~\cite{migdal1941ionization,feinberg1965ionization}).

The sudden change in atomic potential, from Z to Z+1, during $\beta$ decay can lead to secondary shake-off or shake-up processes in which bound electrons are either ejected entirely or excited to higher orbitals, respectively~\cite{Carlson1973}. For the former shake-off process, the energy of the emitted electrons corresponds to the difference in binding energies between $^{32}$P and $^{32}$S. As outer-shell electrons are most likely to be affected, the emitted electrons can carry kinetic energies up to a few tens of eV. The latter shake-up process populates excited atomic states of $^{32}$S; de-excitation of these states produces Auger electrons and characteristic X-rays with energies determined by the $^{32}$S atomic structure.

An analogous source of low-energy electron emission is the Migdal effect, arising from the sudden acceleration of the recoiling $^{32}$S daughter ion and from subsequent nuclear recoils of silicon lattice atoms~\cite{ibe2018migdal,Knapen2021}. Unlike shake-off processes, which result from the change in nuclear charge during the decay, the Migdal effect results from the sudden change in momentum of the recoiling nucleus and can produce ionization electrons with energies in the eV–keV range. A direct observation of the Migdal effect in neutron-induced nuclear recoils of a gaseous target has recently been reported ~\cite{Yi:2026fmf}, and the implications of the Migdal effect for dark matter exclusion limits have been examined in Ref.~\cite{Berghaus:2026kmj}. The effect has not, however, been observed in solid-state systems, and its contribution to the startpoint spectrum therefore remains to be quantified. 

Near the $\beta$ startpoint, bremsstrahlung radiation can be emitted by the slow $\beta$ electron traversing the Coulomb field of the daughter $^{32}$S nucleus. Additionally, as the electron thermalizes within the silicon lattice, it may produce further soft bremsstrahlung through scattering off lattice ions. If these soft photons are fully absorbed within the 0.1~cm$^{3}$ silicon substrate, their energy is recovered in the phonon signal and no spectral distortion results. Photons escaping the substrate would result in distortions to the startpoint spectrum.

Two geometric effects also warrant consideration. First, a fraction of the $\beta$ energy may escape the 0.1~cm thick detector, distorting the higher-energy tail of the startpoint spectrum. While this effect does not impact the startpoint feature itself, it does affect the normalization and shape of the SM background one would wish to fit. Second, a small fraction of $^{32}$P nuclei residing within a thin near-surface layer may produce events in which the daughter ion escapes the substrate, leading to incomplete energy collection and a low-energy tail on the startpoint feature. Given a uniform spatial distribution of neutron-activated $^{32}$P throughout the 0.1~cm$^{3}$ wafer, the surface-to-volume ratio suppresses this contribution at the approximate ppm-level, rendering it subdominant to the LEE background.

Detailed calculations and simulations of these processes and their impact on the $\beta$ startpoint spectrum are beyond the scope of this work. Nevertheless, these higher-order effects must be carefully characterized in order to disentangle a possible sterile neutrino signal from Standard Model spectral features.

\subsection{Systematic Uncertainties}

In Sec.~\ref{sec:form}, we mentioned the systematic uncertainty on the startpoint energy arising from the $Q$-value uncertainty. Though these uncertainties are small with respect to the $\beta$ startpoint energy, increased precision in the $Q$-value and the $\beta$ spectral shape at low $\beta$ energies will become important if precision fits are to be performed. In addition, an accurate determination of the $^{32}$P activity is also relevant for the sterile neutrino sensitivity projections. An important procedural step will be to independently calibrate the $^{32}$P activity within the silicon substrate.

Careful consideration will have to be given to disentangling the dual physics goals of this measurement — searching for sterile neutrinos and calibrating the nuclear recoil response. If defect-induced quenching of nuclear recoils is present, the observed startpoint energy could be shifted to a value below the predicted 49.1~eV by a constant offset related to the energy lost to lattice defect creation. This shift would be degenerate with the effect of a sterile neutrino, which also shifts the startpoint to lower energies. However, the two effects could be distinguished by their spectral signatures: quenching produces a uniform energy scale shift that affects the entire spectrum, whereas sterile neutrino emission produces a distinct secondary spectral feature whose shape and position depend on the neutrino mass and coupling strength. Additionally, performing the measurement with multiple isotopes, for example $^{32}$P and $^{90}$Y, which have different $Q$-values, daughter masses, and consequently different startpoint energies, would provide independent constraints on the quenching factor and break the degeneracy with the predicted sterile neutrino signal. 

We emphasize that either outcome of the quenching measurement is scientifically valuable. If defect-induced quenching is observed, it would imply that current sub-GeV dark matter limits from athermal phonon detectors, which assume equivalent response to nuclear and electronic recoils, are weaker than reported, and previously published exclusion limits would require correction. If no quenching is observed, it validates a key assumption underpinning these searches and strengthens confidence in existing dark matter constraints. In either case, the measurement provides the first direct calibration of the nuclear recoil response in athermal phonon detectors at energies below 100~eV, filling a critical gap in the calibration program for these detectors.

\subsection{Alternative Isotopes and Substrates}

While this work has focused on $^{32}$P in silicon, the $\beta$ startpoint technique is broadly applicable to any $\beta$ decaying isotope that can be embedded in an athermal phonon detector. The choice of isotope is governed by several considerations: isotopes with larger $Q$-values produce higher startpoint energies that are more easily resolved above the detector threshold and the LEE background; isotopes whose daughter nuclei are well-matched in mass to the detector substrate provide the most directly relevant nuclear recoil calibrations; and isotopes that can be produced cleanly via neutron activation of a stable precursor already present in the substrate are preferred. With these criteria in mind, we identify two isotopes of particular interest beyond $^{32}$P.

The first is $^{90}$Y, which can be produced via thermal neutron capture on $^{89}$Y in an analogous fashion to the $^{31}$P$(n,\gamma)^{32}$P process described in Sec.~\ref{sec:exp}. An yttrium-doped silicon substrate could be neutron-activated to produce a well-controlled $^{90}$Y activity. With a $Q$-value of 2278.5~keV, the $^{90}$Y $\beta$ startpoint produces a $^{90}$Zr daughter recoil at 31.0~eV, well within the reach of current detectors. A measurement of the $\beta$ startpoint with $^{90}$Y in a separate detector would provide a powerful systematic cross-check of any sterile neutrino signal or exclusion achieved with $^{32}$P. We note that the yttrium doping would likely have to proceed through ion implantation.

The second isotope of interest is $^{76}$As, a high-$Q$ ($Q = 2960.6$~keV) $\beta$ emitter that produces a $^{76}$Se daughter recoil of 62.0~eV at the $\beta$ startpoint — comfortably above current detector thresholds and comparable to the $^{32}$P recoil energy. It can be produced via thermal neutron capture on $^{75}$As, which is a common $n$-type dopant in germanium. This makes $^{76}$As a natural candidate for measurements in germanium-based athermal phonon detectors. The primary practical challenges of this isotope are its relatively short half-life (26~h), which imposes tighter constraints on the time between neutron activation and the onset of data-taking, and its more complex decay scheme, which includes a significant branching fraction to excited states of $^{76}$Se producing additional gamma rays that must be accounted for in the background model. Despite these complications, $^{76}$As in germanium represents a compelling extension of the $\beta$ startpoint technique to a different substrate and detector technology, and would provide nuclear recoil calibration data directly relevant to the germanium-based dark matter search and CE$\nu$NS detection programs.

\section{Outlook}
\label{sec:outlook}

The measurement proposed here opens a new window on weak nuclear physics at ultralow energies. The $\beta$ startpoint provides simultaneous sensitivity to sterile neutrinos and the nuclear recoil response in solid-state detectors, two of the most pressing questions in particle physics and in sub-GeV dark matter and CE$\nu$NS detection. While practical challenges remain, particularly in source production logistics, background characterization, and the treatment of atomic-scale systematic effects, each of these challenges appears tractable with current or near-term technologies. Initial R\&D will focus on demonstrating neutron activation of phosphorus-doped silicon wafers, characterizing the resulting backgrounds, and performing the first measurement of the $\beta$ startpoint spectrum. We anticipate that this technique will serve as both a discovery tool for sterile neutrinos and an essential calibration method for the next generation of sub-GeV dark matter searches.

\emph{Acknowledgments} --- The authors thank Peter Sorensen for insights on this work. DK thanks Giacomo Marocco for useful discussions. This work was supported by the U.S. Department of Energy (DOE) Office of Science, Office of High Energy Physics under contract DE-AC02-05CH11231.


\pagebreak

\bibliographystyle{apsrev4-2}
\bibliography{references}

\clearpage
\pagebreak

\widetext
\begin{center}
\textbf{\large Supplemental Materials}
\end{center}
\setcounter{equation}{0}
\setcounter{figure}{0}
\setcounter{table}{0}
\setcounter{page}{1}
\makeatletter
\renewcommand{\theequation}{S\arabic{equation}}
\renewcommand{\thefigure}{S\arabic{figure}}
\renewcommand{\thetable}{S\arabic{table}}

\section*{Kinematic Calculation of $\beta$ Startpoint}
\label{sup:A}

The $Q$-value of a $\beta$ decay is defined as the mass difference between the parent isotope $m_{p}$ and the daughter isotope $m_{d}$ with the additional subtraction of the electron mass $m_{e}$. Non-zero neutrino masses, $m_{\nu}$ are considered along with the kinetic energies of the emitted $\beta$ ($T_{\beta}$), emitted neutrino ($T_{\nu}$), and recoiling daughter ion ($T_{d}$). In the Standard Model scenario, we would make the assumption that $m_{\nu} = 0$.

\begin{equation}
\label{1}
    Q = T_{\nu} + m_{\nu} + T_{d} + T_{\beta} ~\text{.}
\end{equation}

We can rewrite the above equation replacing kinetic energies with total energies using the relation $E = T + M$ such that the neutrino mass cancels out. We now define the startpoint of the $\beta$ decay spectra to be the point at which $T_{\beta} \rightarrow 0$, i.e. the emitted $\beta$ has zero kinetic energy. This can be contrasted with the endpoint at which $T_{\nu} \rightarrow 0$, i.e. the emitted neutrino has zero kinetic energy. This gives the equation

\begin{equation}
\label{4}
    Q = E_{\nu} + E_{d} - m_{d} =  E_{\nu} + T_{d}  ~\text{.}
\end{equation}

Since the typical recoiling ion energies in $\beta$ decays have kinetic energies of $O$(100~eV) as compared to their GeV masses, it can safely be assumed to be non-relativistic such that

\begin{equation}
\label{5}
    T_{d} = \frac{p_{d}^{2}}{2m_{d}} ~\text{.}
\end{equation}

We can then substitute Eq.~\ref{5} into Eq.~\ref{4} to get 

\begin{equation}
\label{6}
    Q = E_{\nu} + \frac{p_{d}^{2}}{2m_{d}} ~\text{.}
\end{equation}

We then note that at the startpoint of the $\beta$ decay where $T_{\beta} \rightarrow 0$, $p_{\beta}$ must also approach zero. This implies that the emitted neutrino and recoiling ion are back-to-back such that $p_{d} = p_{\nu} = \sqrt{E_{\nu}^{2} - m_{\nu}^{2}}$. Equation~\ref{6} now becomes

\begin{equation}
\label{7}
    Q = E_{\nu} + \frac{E_{\nu}^{2} - m_{\nu}^{2}}{2m_{d}}  ~\text{.}
\end{equation}

This can be written in powers of $E_{\nu}$ as 

\begin{equation}
\label{8}
    E_{\nu}^{2} + 2m_{d}E_{\nu} - m_{\nu}^{2} - 2m_{d}Q = 0 ~\text{.}
\end{equation}

The solutions for this quadratic are 

\begin{equation}
\label{9}
    E_{\nu} = -m_{d} \pm \sqrt{m_{d}^{2} + m_{\nu}^{2} + 2m_{d}Q} ~\text{.}
\end{equation}

This can then be plugged into Eq.~\ref{4} to find the recoiling ion kinetic energy as a function of neutrino mass, recoiling ion mass, and $Q$-value.

\begin{equation}
\label{10}
    T_{d} = Q - E_{\nu} = Q + m_{d} \mp \sqrt{m_{d}^{2} + m_{\nu}^{2} + 2m_{d}Q}  ~\text{.}
\end{equation}

The physical solution is the one yielding a small positive recoil energy, corresponding to the upper minus sign; we adopt this root throughout. Inserting values for $^{32}$P decays, where $Q=1710.66$~keV, $m_{d}=31.972 \cdot m_N$ where $m_N=0.931494$~GeV, and $m_{\nu} = 0$, we find that $T_{d} = 49.13 $~eV.

We can now calculate the shift in the ion recoil at the $\beta$ startpoint as a function of the neutrino mass by expanding the above relation in powers of $m_{\nu}$

\begin{equation}
\label{11}
    \Delta T_{d} = -\frac{m_{\nu}^{2}}{2\sqrt{m_{d}^{2} + 2m_{d}Q}} + \frac{m_{\nu}^4\sqrt{m_{d}^{2} + 2m_{d}Q}}{8m_{d}^{2}(m_{d}+2Q)^{2}} + O(m_{\nu}^6) + ... 
\end{equation}

\begin{figure}[!t]
	\centering
	\includegraphics[width=0.6 \columnwidth]{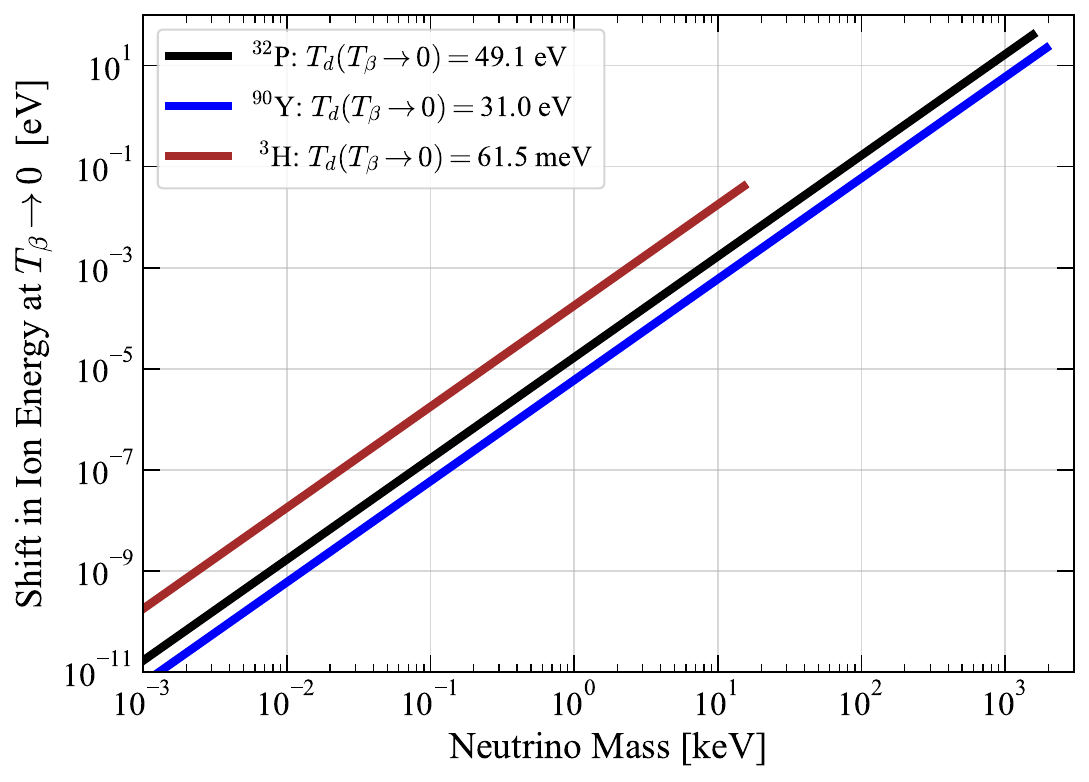}
	\caption{The recoiling daughter ion energy at the $\beta$ startpoint ($T_{\beta} \rightarrow 0$) shifts to lower energies as a function of neutrino mass. Example $\beta$ decaying isotopes shown are $^{32}$P ($Q=1710.66$~keV), $^{90}$Y ($Q=2278.5$~keV), and $^{3}$H ($Q=18.59$~keV).}
	\label{fig:Td_shift}
\end{figure}

The contribution of the $O(m_{\nu}^4)$ term is suppressed by many orders of magnitude, even for MeV-scale neutrino masses, such that terms of order higher than $m_{\nu}^2$ can be safely neglected. Figure~\ref{fig:Td_shift} shows how the daughter ion energy shifts to lower energy with increasing neutrino mass as defined in Eq.~\ref{11}. 

\section*{Impact of neutrino mass on $^{32}$P $\beta$ spectrum}
\label{sup:Aa}

\begin{figure}[!t]
	\centering
	\includegraphics[width=0.6 \columnwidth]{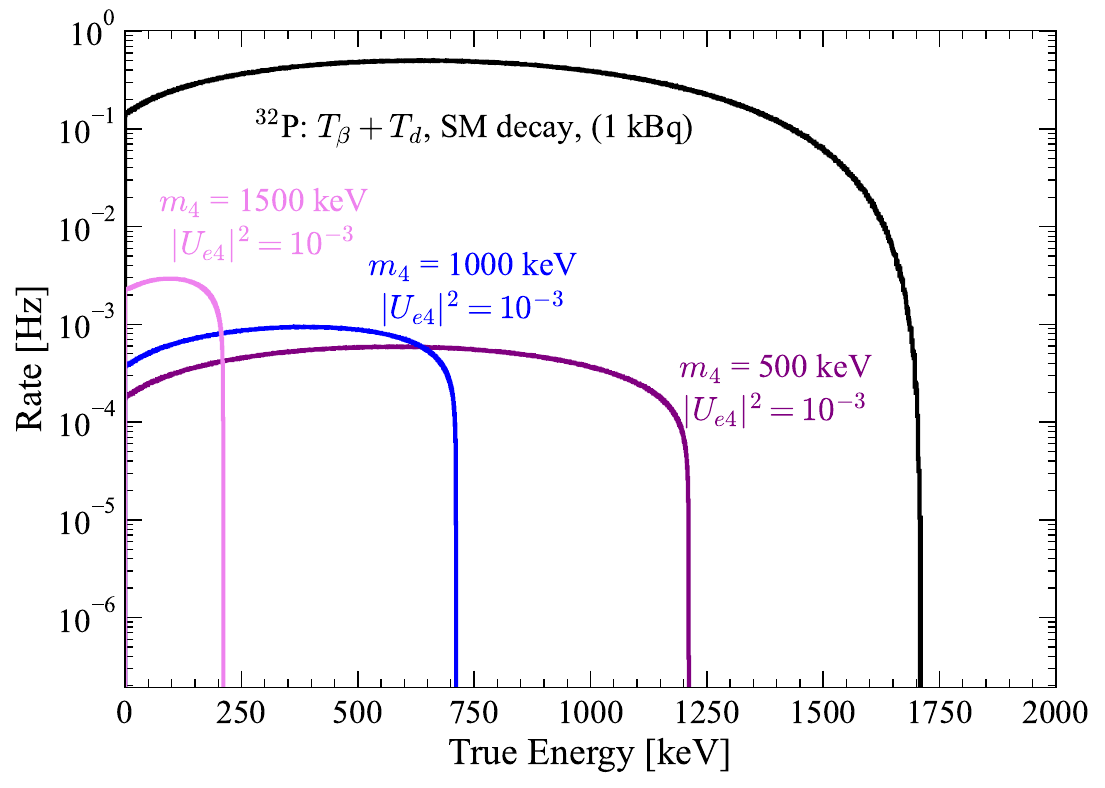}
	\caption{Extension of Fig.~\ref{fig:P32spec}, re-binned and extended out to the $\beta$ endpoint. The $^{32}$P $\beta$ spectrum is scaled assuming the same rates as in Sec.~\ref{sec:Sens}. Also shown are the $\beta$ spectra following the emission of 500~keV, 1000~keV, and 1500~keV sterile neutrinos at a coupling strength $|U_{e4}|^{2}=10^{-3}$.}
	\label{fig:Espec_full}
\end{figure}

Figure~\ref{fig:Espec_full} shows the $^{32}$P $\beta$ spectrum extended out to its endpoint energy of $Q=1710.66$~keV. The spectrum is normalized as in Fig.~\ref{fig:P32spec}, assuming a 1~kBq source and one month of live time, and has been re-binned. The $\beta$ spectra following the emission of a massive neutrino are also shown for $m_4 = 500$~keV~(purple), 1000~keV~(blue), and 1500~keV~(pink). The apparent $\beta$ endpoint shifts down by $m_4$, as expected from Eq.~\ref{eq1}, and the shape of the $\beta$ spectrum is modified according to the phase-space factor of Eq.~\ref{eq5}.

\section*{Neutron Activation Calculations}
\label{sup:B}

\begin{figure}[!t]
	\centering
	\includegraphics[width=0.6 \columnwidth]{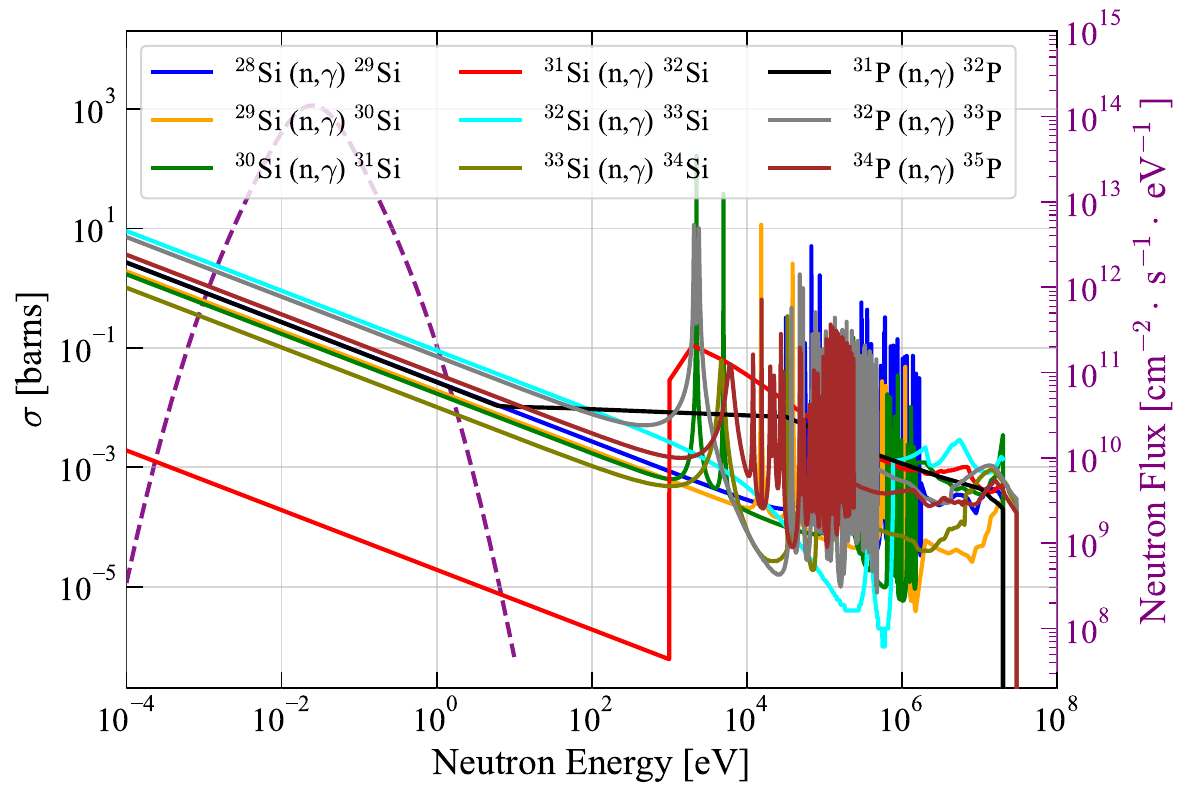}
	\caption{The thermal neutron capture $(n,\gamma)$ cross-sections on silicon and phosphorus isotopes are shown alongside a representative thermal neutron flux from a nuclear reactor facility in purple. This flux and cross-sections were used in the calculations presented in this supplement. }
	\label{fig:flux_ng_xsec}
\end{figure}

We consider neutron capture as the primary mechanism to produce $^{32}$P given a 0.1~cm$^{3}$ silicon wafer containing $^{31}$P nuclei. The silicon wafer is assumed to have natural abundances of pure silicon with dopant densities at the ppm-level. As a benchmark scenario, we examine the required neutron fluences to produce the requisite 1~kBq activity of $^{32}$P considering only $(n,\gamma)$ interactions. The $(n,p)$ interactions are neglected as their cross-sections are significantly smaller at thermal neutron energies. We further assume an idealized thermal neutron source, shown as the purple curve in Fig.~\ref{fig:flux_ng_xsec}, described by a Gaussian with mean energy 0.025~eV and 0.5~eV standard deviation whose total neutron flux is $10^{13}~n/\text{cm}^{2}/$s. 

We first define the production rate as
\begin{equation}
\label{12}
    r = \int \Phi(E)\sigma(E) dE ~,
\end{equation}
where $\Phi$ is the differential neutron flux [$n/\text{cm}^{2}/$s$/\text{eV}$] and $\sigma$ is the $(n,\gamma)$ cross-section. The cross-sections for the $(n,\gamma)$ are shown in Fig.~\ref{fig:flux_ng_xsec} and are taken from ENDF/B-VIII.0~\cite{ENDF} where available and JEFF-4.0~\cite{JEFF} elsewhere ($^{32}$P and $^{33}$P).

We consider that the number of nuclei of a given species within the target doped-silicon can vary in time. Production of one $^{32}$P atom would necessarily lead to fewer $^{31}$P atoms. We further consider that some species may also decay in significant quantities during prolonged neutron exposures. To account for these effects, we construct a series of coupled ordinary differential equations describing the time dependent rates of the silicon and phosphorus isotopes. In this formalism the number of nuclei for a given species is described by $N_{i}$, with corresponding production rate $r_{i}$. Non-stable species are further assigned decay constants, $\lambda_{i}$. The half-lives for $^{31}$Si, $^{32}$Si, $^{33}$Si, $^{32}$P, and $^{33}$P are 2.62~h, 153~y, 6.11~s, 14.3~d, 25.4~d, respectively.
\begin{equation}
\label{13}
\begin{aligned}
        \frac{dN_{Si28}}{dt} &= - r_{Si28}N_{Si28}~, \\
        \frac{dN_{Si29}}{dt} &= r_{Si28}N_{Si28} - r_{Si29}N_{Si29}~, \\
        \frac{dN_{Si30}}{dt} &= r_{Si29}N_{Si29} - r_{Si30}N_{Si30}~, \\
        \frac{dN_{Si31}}{dt} &= r_{Si30}N_{Si30} -(\lambda_{Si31}+r_{Si31})N_{Si31}, \\
        \frac{dN_{P31}}{dt} &= \lambda_{Si31}N_{Si31} - r_{P31}N_{P31}~, \\
        \frac{dN_{Si32}}{dt} &= r_{Si31}N_{Si31} - (\lambda_{Si32}+r_{Si32})N_{Si32}~, \\
        \frac{dN_{P32}}{dt} &= \lambda_{Si32}N_{Si32}
                             +r_{P31}N_{P31}
                             -(\lambda_{P32}+r_{P32})N_{P32}~, \\  
        \frac{dN_{Si33}}{dt} &= r_{Si32}N_{Si32}-(\lambda_{Si33}+r_{Si33})N_{Si33}~, \\
        \frac{dN_{P33}}{dt}  &= \lambda_{Si33}N_{Si33}
                               +r_{P32}N_{P32}
                               -(\lambda_{P33}+r_{P33})N_{P33}~.      
\end{aligned}
\end{equation}

At $t=0$, the number of silicon isotopes is chosen to match natural abundances~\cite{isotopeabunds}: $^{28}$Si (92.2\%), $^{29}$Si (4.7\%), and $^{30}$Si (3.1\%). The number of nuclei is $N_{Si28}=4.6\cdot10^{21}$, $N_{Si29}=2.3\cdot10^{20}$, $N_{Si30}=1.5\cdot10^{21}$. We consider two scenarios for $^{31}$P dopant: (1) no $^{31}$P at $t=0$ (i.e. only natural silicon is present in the wafer) and (2) there are $10^{16}$ $^{31}$P nuclei corresponding to a density of 10$^{17}$~cm$^{-3}$ and a resistivity of $\sim0.1~\Omega\cdot\text{cm}$ (a commercially available density). The remaining isotopes are all assumed to have zero nuclei at $t=0$.

\begin{figure}[!t]
	\centering
	\includegraphics[width=0.6 \columnwidth]{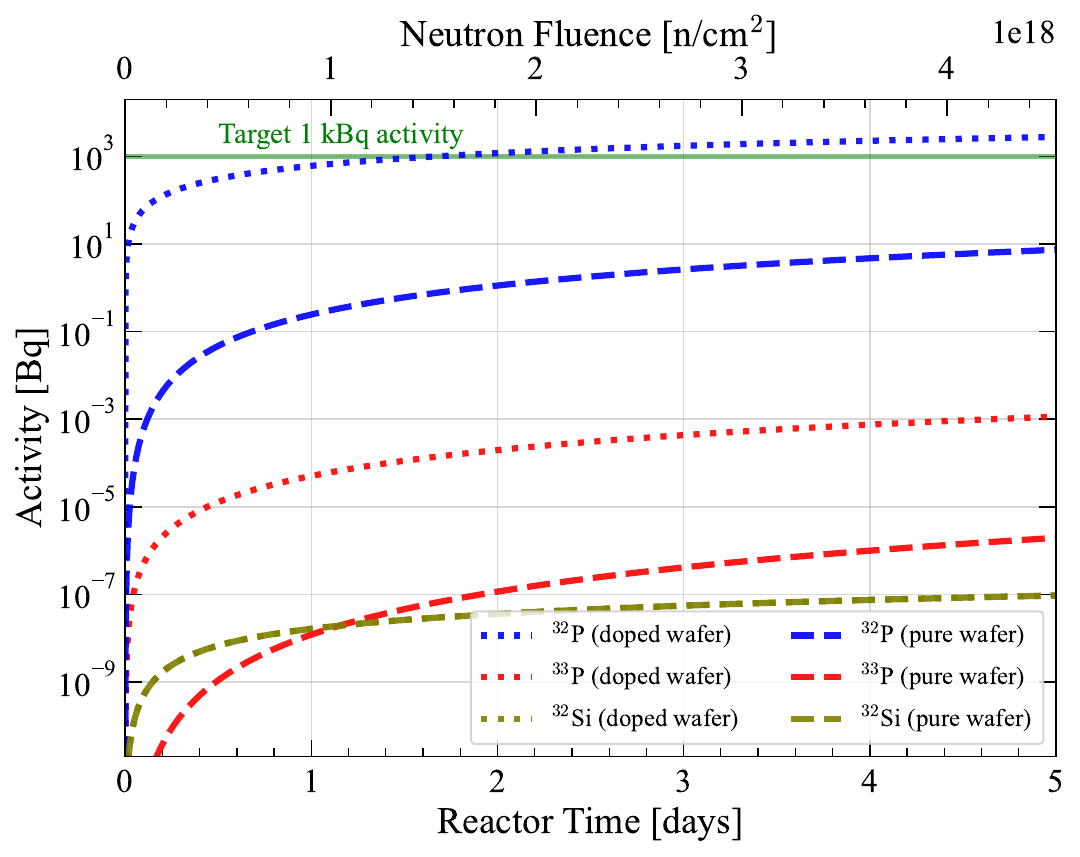}
	\caption{The activity of radioactive isotopes $^{32}$P, $^{33}$P, and $^{32}$Si are shown as blue, red, olive, respectively, assuming $(n,\gamma)$ interactions from an idealized thermal neutron source. The neutron source is described by a Gaussian of mean 0.025~eV and standard deviation 0.5~eV with total flux of $10^{13}~n/\text{cm}^{2}/$s. The target is a 0.1~cm$^{3}$ silicon wafer with either zero dopant (dashed line) or a $^{31}$P dopant density of 10$^{17}$~cm$^{-3}$ (dotted line). Commercially available wafers can have this level of dopant densities. Note the dashed and dotted olive lines are on top of each other.}
	\label{fig:neutronAct}
\end{figure}

Figure~\ref{fig:neutronAct} shows the activity of $^{32}$P as a function of time for the two scenarios described above. The target 1~kBq activity can be reached in just under 2 days of exposure in our idealized thermal neutron reactor when starting with a $^{31}$P-doped silicon wafer. The dotted and dashed lines show the difference between using a doped silicon wafer and a pure silicon wafer. In the latter case, longer exposures and larger neutron fluxes are required.

Also shown in the same figure, as described in Eq.~\ref{13}, is the activity of other long-lived radioactive isotopes, $^{32}$Si and $^{33}$P. The production of these isotopes is a consequence of performing neutron activation on a target that contains $^{31}$Si and $^{32}$P. The activity of these isotopes is greatly suppressed with respect to $^{32}$P. Yet, they have half-lives in excess of $^{32}$P  and so will be present throughout the running of an experiment. In this scheme, one could increase the suppression of the activity of these isotopes by about an order of magnitude by shortening the neutron exposure time: either at the cost of a lower final $^{32}$P activity or by using a higher flux neutron source. The presence of these isotopes could be circumvented by instead embedding radioactive $^{32}$P directly into the silicon wafer.

\end{document}